# Elastic properties of superconducting MAX phases from first principles calculations

I. R. Shein *, A. L. Ivanovskii

Institute of Solid State Chemistry, Ural Branch of the Russian Academy of Sciences, 620990, Ekaterinburg, Russia

**Abstract.** Using first-principles density functional calculations, a systematic study on the elastic properties for all known superconducting MAX phases ($Nb_2SC$, $Nb_2SnC$, $Nb_2AsC$, $Nb_2InC$, $Mo_2GaC$ and $Ti_2InC$) was performed. As a result, the optimized lattice parameters, independent elastic constants, indicators of elastic anisotropy and brittle/ductile behavior as well as the so-called machinability indexis were calculated. We derived also bulk and shear moduli, Young's moduli, and Poisson's ratio for ideal polycrystalline MAX aggregates. The results obtained were discussed in comparison with available theoretical and experimental data and elastic parameters for other layered superconductors.

__________

E-mail: shein@ihim.uran.ru

# 1 Introduction

The group of materials known as nanolaminates or $M_{n+1}AX_n$ (MAX) phases attracted great interest owing to a remarkable combination of properties, which are characteristic both of metals and ceramics. Like ceramics, they are lightweight, elastically rigid, maintain strength to high temperatures *etc.*, whereas like metals, MAX phases are electrically and thermally conductive, plastic and damage tolerant, see reviews [1, 2].

The above metallic-like nature of MAX phases allows us to consider these materials as potential superconductors (SCs). Indeed, at present among ~ 60 synthesized $M_{n+1}AX_n$ phases [2], six low-temperature SC have been discovered: $Mo_2GaC$ ($T_C$ ~ 4K [3]), $Nb_2SC$ ($T_C$ ~ 5K [4]), $Nb_2SnC$ ($T_C$ ~ 7.8K [5]), $Nb_2AsC$ ($T_C$ ~ 2K [6]), $Ti_2InC$ ($T_C$ ~ 3K [7]) and $Nb_2InC$ ($T_C$ ~ 7.5K [8]).

These materials possess a layered hexagonal structure (where blocks of transition metal carbides [MC] formed by edge-shared $M_6C$ octahedra are sandwiched with A atomic sheets) and thus are similar to other groups of layered SCs such as high-$T_C$ oxides, $MgB_2$, graphite intercalation compounds (GICs), borocarbides ($YNi_2B_2C$, $LuNi_2B_2C$), carbide (nitride) halides ($\beta$-HfNCl, $Y_2C_2I_2$) or recently discovered FeAs-SCs, which attract now much attention as *anisotropic* SCs, see [9-12].

Elastic properties of crystals are closely related to many fundamental properties, such as equation of state, thermal expansion, melting point, and some others. From elastic parameters, valuable information can be obtained about intra-atomic bonding, structural stability *etc*. Furthermore, elastic properties are of great interest for the material science of SCs; for example, elastic constants can be linked to such important physical parameters of SCs as the Debye temperature $\Theta_D$ and the electron-phonon coupling constant $\lambda$. Besides, mechanical properties are important for technology and advanced applications of superconducting materials [13-15].

In this paper, we present a detailed theoretical study on the elastic properties *for all known superconducting MAX phases*: $Nb_2SC$, $Nb_2SnC$, $Nb_2AsC$, $Nb_2InC$, $Mo_2GaC$ and $Ti_2InC$.

# 2 Computational method

The considered superconducting $M_2XC$ ($M$ = Mo, Nb and Ti; $X$ = In, Ga, Sn, S and As) phases possess the hexagonal structure with space group $P6_3/mmc$ (No. 194), where blocks of transition metal carbides [$MC$] (formed by edge-shared $M_6C$ octahedra) are sandwiched with $X$ atomic sheets. The Wyckoff positions of atoms in $M_2XC$ are – carbon: $2a$ (0, 0, 0), $X$: $2d$ (1/3, 2/3, 3/4), and $M$ atoms: $4f$ (1/3, 2/3, $z_M$). The structure is defined by the lattice parameters $a$ and $c$, and the internal parameter $z_M$, see [1, 2].

All calculations were based on the density-functional theory as implemented in the Vienna *ab initio* simulation package (VASP) in projector augmented waves (PAW) formalism [16, 17]. Exchange and correlation were described by a non-local correction for LDA in the form of GGA [18]. The kinetic energy cutoff of 500 eV and k-mesh of 16×16×6 were used. The geometry optimization was performed with the force cutoff of 2 meV/Å.

## 3 Results and Discussion

As the first step, the equilibrium lattice constants (*a* and *c*) for all the above $M_2XC$ phases were calculated with full structural optimization including internal parameters $z_M$. The results obtained (Table 1) are in reasonable agreement with the available theoretical and experimental data.

Next, the values of five independent elastic constants for hexagonal crystals ($C_{11}$, $C_{12}$, $C_{13}$, $C_{33}$ and $C_{44}$) were calculated (Table 2), which allowed us to obtain the bulk *B* and shear *G* moduli. Usually, for such calculations two main approximations are used, namely the Voigt (V) [19] and Reuss (R) [20] schemes.

Thus, in terms of the Voigt approximation, these moduli are:
$B_V = (1/9) \{2(C_{11} + C_{12}) + 4C_{13} + C_{33}\}$,
$G_V = (1/30) \{C_{11} + C_{12} + 2C_{33} - 4C_{13} + 12C_{44} + 12C_{66}\}$
In terms of the Reuss approximation:
$B_R = \{(C_{11} + C_{12})C_{33} - 2C_{12}^2\}/(C_{11} + C_{12} + 2C_{33} - 4C_{13})$,
$G_R = (5/2) \{[(C_{11} + C_{12})C_{33} - 2C_{12}^2] C_{55}C_{66}\}/\{3B_V C_{55}C_{66} + [(C_{11} + C_{12})C_{33} - 2C_{12}^2]^2(C_{55} + C_{66})\}$

The above elastic moduli were obtained from first-principles calculations for monocrystals. Meanwhile, the majority of synthesized and experimentally examined $M_2XC$ phases were prepared and investigated as polycrystalline ceramics, *i.e.* in the form of aggregated mixtures of microcrystallites with random orientation. For estimations of elastic parameters for polycrystalline materials, the Voigt-Reuss-Hill (VRH) approximation [21] is widely used, where the actual effective moduli for polycrystals are expressed as the arithmetic mean of the two above mentioned limits - Voigt and Reuss: $B = 1/2(B_V + B_R)$ and $G = 1/2(G_V + G_R)$, and averaged compressibility in the form $\beta = 1/B$.

Further, the calculated bulk moduli *B* and shear moduli *G* allow us to obtain the Young's moduli *Y*, the Poisson's ratio *v* and the Lame's constants $\mu$, $\lambda$ as:
$Y = 9BG/(3B + G)$
$v = (3B - 2G)/\{2(3B + G)\}$,
$\mu = Y/2(1 + v)$, $\lambda = vY/\{2(1 + v)(1 - 2v)\}$,

The above elastic parameters presented in Tables 2 and 3 allow us to make the following conclusions:

(i). The $C_{ij}$ constants for all $M_2XC$ phases are positive and satisfy the generalized criteria [22] for mechanically stable crystals: $C_{44} > 0$, $C_{11} > |C_{12}|$, and $(C_{11} + C_{12}) C_{33} > 2C_{13}^2$.

(ii). Among superconducting $M_2XC$ phases, $Mo_2GaC$ is the phase with the largest bulk modulus (~ 249 GPa), while $Ti_2InC$ has the smallest $B$ ~ 125 GPa; in turn, the smallest and largest shear moduli are ~ 67 GPa ($Nb_2SnC$) and ~ 96 GPa ($Mo_2GaC$). Thus, for all of the considered phases $B > G$; this implies that the parameter limiting the mechanical stability of these materials is the shear modulus. Note also that the trends in bulk moduli for $M_2XC$ phases ($B(Mo_2GaC) > B(Nb_2XC) \sim B(Ti_2InC)$, Table 3) repeat the same for binary cubic carbides ($B(MoC) = 337$ GPa $> B(NbC) = 301$ GPa $\sim B(TiC) = 242$ GPa [27]). This allows us to conclude that [$MC$] blocks of $M_2XC$ phases are responsible for these changes.

(iii). According to Pugh's criteria [28], a material should behave in a ductile manner if $G/B < 0.5$, otherwise it should be brittle. In our case, $G/B(Mo_2GaC) \sim G/B(Nb_2SC) = 0.40 < G/B(Nb_2InC) = 0.44 < G/B(Nb_2AsC) = 0.50 < G/B(Nb_2SnC) = 0.59 < G/B(Ti_2InC) = 0.66$; i.e. according to this indicator $Mo_2GaC$ and $Nb_2InC$ will behave in a ductile manner, $Nb_2AsC$ will lie on brittle/ductile border, whereas $Nb_2SnC$ and $Ti_2InC$ will demonstrate brittleness. An additional argument for the variation in the brittle/ductile behavior of $M_2XC$ phases follows from the calculated Poisson's ratio ν, Table 3. Indeed, for brittle materials these values are small enough, whereas for ductile metallic materials ν is typically 0.33 [29]. In our case, for ductile $Mo_2GaC$ $v \sim 0.33$ is close to this limit and is the highest value among those for other superconducting $M_2XC$ phases, Table 3.

(iv). Another indicator of useful performance is the so-called machinability index $\mu_M = B/C_{44}$ [30]. In our case, $\mu_M$ ($Nb_2AsC$) $< \mu_M(Ti_2InC) = 2.17 < \mu_M(Mo_2GaC) = 2.44 \sim \mu_M(Nb_2SC) = 2.50 < \mu_M(Nb_2SnC) = 2.91 < \mu_M(Nb_2InC) = 3.19$; i.e. among superconducting phases $Nb_2InC$ exhibits the largest machinability index, which, nevertheless, is much smaller than the same for isostructural MAX phases with highest machinability: $W_2SnC$ ($\mu_M = 33.3$) and $Mo_2PbC$ ($\mu_M = 15.8$) [23].

(v). The Young's modulus is defined as the ratio between stress and strain and is used to provide a measure of stiffness, i.e., the larger is the value of $Y$, the stiffer is the material. In our case $Y(Nb_2AsC) > Y(Mo_2GaC) > Y(Nb_2SC) > Y(Nb_2InC) > Y(Ti_2InC) > Y(Nb_2SnC)$, Table 3.

(vi) Elastic anisotropy of crystals is a very important factor for material science of superconductors since it correlates with the possibility of appearance of microcracks in these materials [31, 32]. There are different ways to estimate elastic anisotropy theoretically. For example, the so-called shear anisotropy ratio $A = 2C_{44}/(C_{11} - C_{12})$ is widely used for this purpose. The factor $A_1 = 4C_{44}/(C_{11} + C_{33} - 2C_{13})$ is also used as a measure of the degree of anisotropy for the {100} shear planes between ‹011› and ‹010› directions. Our estimations (Table 3) demonstrate that all the examined phases are anisotropic, but In-containing phases $Ti_2InC$, $Nb_2InC$ and $Nb_2AsC$ exhibit the maximal deviation from $A, A_1 = 1$.

(vii). The ratio between linear compressibility coefficients $k_c/k_a$ of hexagonal crystals may be obtained from the index $f = (C_{11} + C_{12} - 2C_{13})/(C_{33} - C_{13})$ [33]. Our data show that (i) $Nb_2SnC$ lies close to the isotropic limit $f = 1$; (ii) for $Nb_2SC$, $Nb_2AsC$, $Nb_2InC$ and $Mo_2GaC$ $f < 1$, which means that the compressibility for these crystals along the $c$ axis is smaller than along the $a$ axis;

but (iii) for $Ti_2InC$ a reversed situation takes place: $f > 0$, *i.e.* for this material the *c* direction is softer.

**4 Conclusions**

Based on first principles calculations, a systematic study on the elastic properties for all known superconducting $M_2XC$ phases ($Nb_2SC$, $Nb_2SnC$, $Nb_2AsC$, $Nb_2InC$, $Mo_2GaC$ and $Ti_2InC$) was performed.

The evaluated elastic parameters allow us to conclude that all the superconducting $M_2XC$ phases are mechanically stable; the parameter limiting their mechanical stability is the shear modulus. Our data reveal the intervals of elastic parameters, which are typical of superconducting $M_2XC$ phases. So, the maximal and minimal values of bulk moduli are 249 GPa ($Mo_2GaC$) and 125 PGa ($Ti_2InC$); the shear moduli vary from 117 GPa (for $Nb_2AsC$) to 67 GPa (for $Nb_2SnC$).

These anisotropic superconductors include both ductile ($Mo_2GaC$, $Nb_2InC$) and brittle ($Nb_2SnC$, $Ti_2InC$) materials; their machinability index varies in a rather narrow interval from 2.17 ($Nb_2AsC$, $Ti_2InC$) to 3.19 ($Nb_2InC$), which is much smaller than the same for isostructural MAX phases [23].

Finally, let us note that the elastic parameters of $M_2XC$ superconductors are higher than for example for layered FeAs SCs, which are relatively soft materials ($B < 100$ GPa) [34, 35], but are comparable with the same for some other layered SCs (such as YBCO, $MgB_2$, borocarbides, carbide halides of rare-earth metals *etc*.), for which their bulk moduli do not exceed $B \leq 200$ GPa, see [34].

**Acknowledgments**
Financial support from the RFBR (Grant 09-03-00946-a) is gratefully acknowledged.

**Table 1.** Calculated lattice parameters ($a$, $c$, in Å), ratio $a/c$ and internal parameters $z_M$ for superconducting MAX phases in comparison with available data.

| phase | **Mo$_2$GaC** | **Nb$_2$InC** | **Nb$_2$AsC** | **Nb$_2$SnC** | **Nb$_2$SC** | **Ti$_2$InC** |
|---|---|---|---|---|---|---|
| $a$ | 3.0680 (3.01 [a]; 3.084 [d]) | 3.1933 (3.17 [a]; 3.196 [d]) | 3.3442 (3.31 [a]; 3.339 [d]) | 3.2771 (3.241 [a]; 3.273 [d]) | 3.3204 (3.294 [b]; 3.312 [d]) | 3.1485 (3.13 [a]; 3.132 [c]; 3.148 [d]) |
| $c$ | 13.1533 (13.18 [a]; 13.16 [d]) | 14.4952 (14.37 [a]; 14.47 [d]) | 12.048 (11.9 [a]; 12.00 [d]) | 13.9029 (13.802 [a]; 13.85 [d]) | 11.7093 (11.553 [b]; 11.65 [d]) | 14.2071 (14.06 [a,b]; 14.20 [d]) |
| $c/a$ | 4.287 | 4.539 | 3.603 | 4.242 | 3.527 | 4.512 |
| $z_M$ | 0.0889 | 0.0819 | 0.0944 | 0.0828 | 0.0952 | 0.0780 |

* available experimental and theoretical data are given in parentheses
[a] Reference [1]; [b] Reference [4]; [c] Reference [7]; [d] Reference [23];

**Table 2.** Calculated elastic constants ($C_{ij}$, in GPa) for superconducting MAX phases in comparison with available data.

| phase | Mo$_2$GaC | Nb$_2$InC | Nb$_2$AsC | Nb$_2$SnC | Nb$_2$SC | Ti$_2$InC |
|---|---|---|---|---|---|---|
| $C_{11}$ | 306.4 (294 [a]) | 291.3 (291 [a]) | 325.3 (327 [a]) | 254.8 (252 [a]; 341 [b]; 311 [c]) | 303.6 (309 [a]) | 282.6 (282 [a]; 273 [d]) |
| $C_{12}$ | 105.1 (98 [a]) | 77.4 (76 [a]) | 113.7 (108 [a]) | 100.8 (96 [a]; 105 [b]; 96 [c]) | 116.9 (106 [a]) | 70.2 (65 [a]; 63 [d]) |
| $C_{13}$ | 169.0 (160 [a]) | 117.6 (108 [a]) | 161.3 (155 [a]) | 120.0 (129 [a]; 169 [b]; 97 [c]) | 155.1 (159 [a]) | 54.9 (55 [a]; 50 [d]) |
| $C_{33}$ | 311.2 (298 [a]) | 288.7 (267 [a]) | 325.8 (347 [a]) | 243.0 (244 [a]; 320 [b]; 306 [c]) | 315.7 (310 [a]) | 232.9 (240 [a]; 232 [d]) |
| $C_{44}$ | 102.1 (86 [a]) | 57.1 (102 [a]) | 150.0 (162 [a]) | 58.9 (99 [a]; 183 [b]; 119 [c]) | 88.1 (118 [a]) | 57.6 (86 [a]; 87 [d]) |

\* available theoretical data are given in parentheses
[a] Reference [23]; [b] Reference [24]; [c] Reference [25]; [d] Reference [26];

**Table 3.** Calculated bulk moduli ($B$, in GPa), compressibility ($β$, in GPa$^{-1}$), shear moduli ($G$, in GPa), Young's moduli ($Y$, in GPa), Poisson's ratio ($v$), Lame's constants ($μ, λ$, in GPa), elastic anisotropy parameters ($A, A_1$ and $f$, see the text) for superconducting MAX phases.

| phase | Mo$_2$GaC | Nb$_2$InC | Nb$_2$AsC | Nb$_2$SnC | Nb$_2$SC | Ti$_2$InC |
|---|---|---|---|---|---|---|
| $B$ | 248.6 | 182.4 | 234.3 | 171.1 | 220.9 | 124.7 |
| $β$ | 0.004022 | 0.005482 | 0.004269 | 0.005846 | 0.004527 | 0.008019 |
| $G$ | 95.7 | 79.6 | 117.1 | 66.8 | 88.6 | 81.7 |
| $Y$ | 254.4 | 208.5 | 301.0 | 177.3 | 234.5 | 201.1 |
| $v$ | 0.3294 | 0.3095 | 0.2858 | 0.3272 | 0.3230 | 0.2312 |
| $λ$ | 184.8 | 129.3 | 156.2 | 126.5 | 161.8 | 70.3 |
| $μ$ | 95.7 | 79.6 | 117.1 | 66.8 | 88.6 | 81.7 |
| $A$ | 1.01 | 0.53 | 1.42 | 0.77 | 0.94 | 0.54 |
| $A_1$ | 1.46 | 0.66 | 1.83 | 0.91 | 1.14 | 0.57 |
| $f$ | 0.52 | 0.78 | 0.71 | 0.94 | 0.69 | 1.37 |